\documentclass[preprint,showpacs,preprintnumbers,amsmath,amssymb]{revtex4}

\usepackage{epsfig}
\usepackage{amsmath}
\usepackage{graphicx}
\usepackage{dcolumn}
\usepackage{bm}

\begin{document}

\preprint{{\it  submitted to Physical Review Letters}}

\title{Shock Waves in Nanomechanical Resonators} 

\author{Florian W. Beil}
\address{%
Center for NanoScience and Sektion Physik,  Ludwigs-Maximilians-Universit\"at-M\"unchen,
 Geschwister-Scholl-Platz 1, 80539 M\"unchen, Germany.
 }%
\author{Achim Wixforth}
\address{%
Lehrstuhl f\"ur Experimentalphysik I, Universit\"at Augsburg, Universit\"atsstra{\ss}e 1, D-86135
Augsburg, Germany.
}
\author{Werner Wegscheider}
\address{
Institut f\"ur Angewandte und Experimentelle Physik, Universit\"at
Regensburg, 93040 Regensburg, Germany.
}%
\author{Dieter Schuh}
\address{
Institut f\"ur Angewandte und Experimentelle Physik, Universit\"at
Regensburg, 93040 Regensburg, Germany.
}%
\author{Max Bichler}
\address{
Walter Schottky Institut, Am Coloumbwall 3, 85748 Garching,
Germany.
}%
\author{Robert H. Blick}
\email{blick@engr.wisc.edu}
\address{%
Electrical and Computer Engineering,
University of Wisconsin-Madison, 1415 Engineering Drive, Madison
WI 53706, USA.
}

\date{\today}

\begin{abstract}
{\bf 
The dream of every surfer is an extremely steep wave propagating at the highest speed possible. The best 
waves for this would 
be shock waves, but are very hard to surf. In the nanoscopic world the same is true:  the surfers in this case are electrons
riding through nanomechanical devices on acoustic waves~\cite{achim2}. Naturally, this has a broad 
range of applications in sensor technology and for communication electronics for which the combination of an electronic and a mechanical 
degree of freedom is essential. But this is also of interest for fundamental aspects of 
nano-electromechanical systems (NEMS), when it comes to quantum limited displacement detection~\cite{cleland} 
and the control of phonon number states~\cite{schwab}. 
Here, we study the formation of shock waves in a NEMS resonator with an embedded two-dimensional electron gas using
surface acoustic waves. The mechanical displacement of the nano-resonator is read out via the induced acoustoelectric current. 
Applying acoustical standing waves we are able to determine the anomalous acoustocurrent. This current is only found in the
regime of shock wave formation. We ontain very good agreement with model calculations.  
}
\end{abstract}
\maketitle
  Over the past two decades a multitude of work focused on reducing the dimensionality of electronic systems from three to zero dimensions, 
leading to extremely interesting physics~\cite{textbook_John}. Only recently the availability of even more intricate nano-structuring techniques made it possible to morph the dimensionality of the phonon systems, i.e. to construct phonon cavities with nano-scale dimensions.  An example of this 
are freely suspended two-dimesional electron systems (2DEG), for which the phonon modes can be tailored~\cite{blick}.
The combination of low-dimensional electron gases and nanomechanical systems is interesting for several reasons:  the most obvious
being the ability to study intricate effects of the electron-phonon interaction. In other words dissipation in the limit of single electrons 
interacting with discrete phonon modes~\cite{eva_prl}. This is revealed in the formation of 
van Hove-singularities in the phonon density of states for low-dimensional phonon systems~\cite{brandes}. Furthermore, suspended 2DEGs, quantum 
wires, and quantum dots, are the perfect tools for studying quantum electromechanical (QEM) effects~\cite{blencowe} and might further 
improve measurements for the current standard~\cite{pepper}. 

The main limitation encountered so far is the fact that the phonon system, in contrast to the electronic one, could not be actuated directly.
Nevertheless, the 
method of choice for generating such an acoustic actuation is readily available:  surface acoustic waves (SAW) are a proven tool for investigating
non-suspended 2DEGs already~\cite{achim}. Hence, we set out to combine low-dimensional electron systems embedded in a nanomechanical resonator with 
SAW generators, such as interdigitated transducers (IDTs). In early work we demonstrated how to achieve acoustical coupling of an ordinary
nanomechanical resonator (suspended beam without electron gas) to SAWs~\cite{beil_physE1} and were able to show how acoustical standing wave patterns
can be detected in such a resonator~\cite{beil_physE2}. Recently, we pushed the resolution of these measurements and were able to resolve single defect
relaxtion~\cite{Beil:2} and phonon population inversion in a nano-resonator~\cite{flo_apl}.

In this Letter we present acoustical excitation of a nano-resonator with an integrated 2DEG by SAW, leading to the formation of shock waves in the
nanomechanical system. Shock wave formation is probed with the help of the so-called anomalous
acoustoelectric effect, one of the manifestations of the electron-phonon interaction. 
Acoustoelectric effects have up to now been mainly 
studied in quantum wells, where the mechanical properties are determined by the bulk~\cite{achim,Sasha:1}. 
It is apparent that the investigation 
of acoustoelectric effects in low dimensional mechanical systems is also interesting for improving current 
standards~\cite{pepper,Shilton:1,Talyanskii:1} and elucidate effects like the anomalous acoustoelectric effect observed in thin films itself~\cite{Ilisavskii:1}. The reasoning 
is straight forward:  suspending low-dimensional electron systems leads to giant displacements as compared to SAWs on bulk, hence, the interaction
of electrons and acoustical phonons is extremely enhanced. 

In the following we will describe how SAWs mechanically excite the suspended electronic specimen and how acoustic currents are observed. The measured current consists of two components, of which one is an anomalous current. This current directly traces the shock wave 
formation in the nano-resonator. As shock wave we define the transition from a linear response such as a sinusoidal excitation towards nonlinear wave forms
of the nano-resonator's displacement. 

{\it Methods--}
The processing follows standard techniques for suspended 2DEGs~\cite{Eva:1,Eva:2}, with the difference that we integrate interdigitated transducers (IDTs) for SAW generation on the sample. Starting with an AlGaAs-heterostructure containing a 2DEG grown above a sacrificial layer (300 nm thickess), the freely suspended electronic system is defined by successive lithography steps. The lateral structures of the suspended 2DEGs and the IDTs are then defined via electron beam lithography. In a further step, anisotropic reactive ion etching (RIE) is applied to mill out the lithographically defined structure. Finally, the sacrificial layer is removed in an isotropic wet etch step with hydrofluoric acid. 
In Fig.~\ref{figure0} a scanning electron micrograph of the sample is shown. The inset presents the suspended beam of length
$L$~=1.2~$\mu$m, width $w$~=~300~nm, and height $h$~=~200~nm used in the experiments. 
As seen the suspended sample is placed between two IDTs forming an acoustic delay line. In the following the two IDTs are excited with
continuous waves from two synthesizers. 

For the first set of experiments we generate the acoustical waves in one transducer, while
for shock wave probing we couple both IDTs to generate an acoustical standing wave pattern and then trace the induced direct current.
The transducers generate a coherent acoustic sound wave via the inverse piezo effect, at the lithographically defined center
frequency $f_{\rm saw}$. This corresponds to a SAW wavelength $\lambda_{\rm saw}$ = 7~$\mu$m resp.
9~$\mu$m. The SAW frequency and wavelength are connected via $f_{\rm saw}~=~v_{\rm saw}/\lambda_{\rm saw}$, where $v_{\rm
saw}$~=~2865 m/s is the surface wave velocity on GaAs in the [011]-direction. It has to be noted that no acoustic power is
reflected at the etch boundaries produced by the anisotropic RIE step, due to the small height compared to the SAW wavelength. 
In order to measure the acoustoelectric effects in the 2DEG in a two point fashion, we either employ a lock-in technique (as demonstrated in
Fig.~\ref{figure0}), or measure the direct current driven through the 2DEG with an optional DC bias voltage. Thus
both the resistance of the beam and the direct current can be measured in dependence of acoustic wave excitation.

{\it Experiment--}
For determining the quality of the suspended electron system we first took magnetoresistance traces of  the freely suspended 2DEG at 100~mK (Fig. 2(a)). The longitudinal resistance exhibits $1/B$ periodic Shubnikov-de Haas oscillations,
from which we calculate a carrier density of 6.56 $\times$10$^{15}$ m$^{-2}$ with a mobility of $\mu$~=~3057~cm$^{2}$ /(Vs). The peak in the resistance 
at low magnetic field (cf. inset of Fig.~2(a)) is due to coherent back scattering effects in the suspended
2DEG, as discussed elsewhere~\cite{Eva:1}. A backgate voltage $V_{\rm bg}$ is applied to alter the suspended 2DEG's resistance.
In Fig.~2(b) we show the effect of acoustic excitation by the left and right IDT firing at the suspended beam. 
Scanning the radio frequency signal applied at the left IDT results in a modulation of
the current which can be passed through the sample.
A notable current modulation is only observed when $f$ is approximately  
$f_{\rm saw}$ and a maximal acoustic power is generated. 

In detail, the induced acoustoelectric current is depicted when either the left or the right IDT are driven by one of the synthesizers. While the 
left transducer generates a forward current the right one reverses the current direction and pumps electrons backwards.  These traces are recorded
under increasing RF power (0 dBm ... +10 dBm). 
The dependence of the SAW induced current on IDT-frequency $f$, SAW power and propagation direction are consistent
with previous measurements of the acoustoelectrical current in non-suspended 2DEGs~\cite{Shilton:1} and quantum point contacts~\cite{Talyanskii:1}. 
The amplitude of the normal acoustoelectric current is given by $I_n = n e f_{\rm saw}$, where $n$ is the integer number of electrons transferred:  this suggests that during each SAW cycle $\sim 380$ electrons are transferred through the suspended 2DEG. 

In addition to this conventional acoustoelectric current a peculiar feature appears in Fig.~2(b) at the left 
edge of the IDTs' main transmission centered at 350 MHz:  the acoustic current in the shoulder of the current peak is caused
by an additional current that is {\it not} changing sign with SAW direction. 
This so-called {\it anomalous acoustoelectric} current $I_{\rm an}$ was first observed in thin manganite films~\cite{Ilisavskii:1} and was
explained by the elastic deformation of the thin film by the SAW. The resulting unidirectional current ($I_{\rm an}$) is found to be even in the SAW wave
vector $k_{\rm saw}$, and thus invariant under inversion of SAW direction. The total acoustoelectric current then is given as a
superposition of the normal and anomalous components $I_{\rm n} + I_{\rm an}$. 

It is this anomalous acoustoelectric current, which allows us to probe the deformation of the nanomechanical device directly. 
In order to read out the unidirectional current we conducted acoustic standing wave experiments:  a standing wave
is formed by applying phase locked RF-signals to both IDTs, left and right of the sample. Shifting the relative phase $\phi$ of the driving
signal at one IDT with respect to the other results in a lateral shift of the standing wave pattern. If a perfect standing wave is formed the total wave
vector equals to zero and the only contribution to the measured acoustoelectric current is the anomalous component $I_{\rm an}$. 
{\it In other words the normal acoustoelectric current depends only on the propagating part, whereas the anomalous current depends on the mechanical deformation induced in the suspended 2DEG.} Thus we have a direct relation between the mechanical deformation and the relative phase $\phi$, which allows us
to map the mechanical mode of the suspended beam. 

In Fig.~3(a) the standing wave pattern of the anomalous current is shown:  evidently the initial trace with moderate power levels applied to the IDTs
is a fundamental mode of the suspended nanomechanical device, forming a sinusoidal-trace. Increasing the acoustic power leads to a larger current, i.e.
the trace is increased towards more negative currents.  Evidently, the shape of the standing wave pattern is changed once the relative  phase $\phi$ of the two
synthesizers is altered. The deviation from the sinusoidal-waveform shows up as a pronounced peak around $\phi = 180$~deg. 
This is the transition from linear
to nonlinear response, where the steepening of the trace indicates {\it shock wave formation}. In other words, all of the energy of the 
SAW is compressed around  $\phi = 180$~deg, resulting in this nonlinear wave form. By varying $\phi$, we shift the maxima and 
minima of the acoustic wave through the resonator, thus mapping the deformation of the resonator directly in the current response. 
In Fig.~3(b) the standing wave pattern of the anomalous current is shown for even larger acoustic excitation power levels:  the shock wave form now
develops extremely high current levels around  $\phi = 0$~deg and  $360$~deg.  
A pronounced peak at $\phi$ = 180~deg is accompanied by two apparent shoulders to its left and right. For the highest power levels shown in the figure, the 'wave form' returns to one resembling a sinusoidal again. We interpret this
as the final transition into a higher order mode.

{\it Theory --}
In Fig.~4(a) we show the full power dependence of the acoustic current while altering $\phi$ from smallest to largest driving powers at the IDTs. 
As seen the ground mode shows a slight sinusoidal-modulation, which deforms into a shock wave, 
indicated by the sharp peak and the two shoulders discussed above at intermediate power levels. 
A further inrease in power switches the system to the next higher mode, which again shows a sinusoidal wave form response.
Evidently, this series of traces depicts a transition from the first fundamental mode to the next higher mode with a larger current
amplitude.  

To model this effect we calculated the anomalous acoustoelectric current in the suspended 2DEG, induced by altering the 
phase shift of the acoustic standing wave, following Illisavskii {\it et al.}~\cite{Ilisavskii:1}. 
When an acoustic wave interacts with a thin film, the induced mechanical strain modulates the local 
conductivity, which induces an  acoustoelectric current density $j_{\rm }$ per unit length
\begin{equation}
j_{\rm }(z)=\frac{a \ \omega}{2 \pi}\int_{0}^{\frac{2
\pi}{\omega}} \sigma_{zz}(z,t) E_{z}(z,t) \ dt, \label{Equ1}
\end{equation}
where $\omega$ is the SAW frequency, $a$ the thickness of the resonator, $\sigma_{zz}$ is the component of the conductivity tensor
along the $z$-axes (the [110] direction, cf. Fig.~\ref{figure0}), and $E_{z}$ is the electric field along $z$. Eq.~(\ref{Equ1})
neglects any dependence on the $y$-coordinate, which is justified due to the 'thinness' of the sample of only 200~nm, compared to its
length of $\sim$1~$\mu$m. The SAW induced strain in the beam $S_{ij}$ will modulate the conductivity following
\begin{eqnarray}
\sigma_{zz}(z,t)& = & \sigma_0 (\Pi_{zzzz} S_{zz}(z,t) +
\Pi_{zzyy} S_{yy}(z,t) + \nonumber\\
& & \Pi_{zzzy} S_{yz}(z,t)), \label{Equ2}
\end{eqnarray}
where $\sigma_0$ is the unperturbed conductivity, and the tensor $\Pi_{ijkl}$ describes the effect of strain $S_{jk}$ on the conductivity $\sigma_{ik}$ and evaluates
to $\Pi_{ijkl}=\delta \sigma_{ij} / \delta S_{jk}$. To solve Eq.~(\ref{Equ1}) in combination with Eq.~(\ref{Equ2}) we have to
determine the SAW induced strain in the beam and relate it to $E_{z}$. If two counter propagating SAWs form a standing wave
pattern, the motion of the clamping points will induce stress in the beam, extending from $0$ to $L$ in the $z$-direction, which
is calculated to~\cite{Beil:2}
\begin{eqnarray}
S_{zz}(z,t)& = & \frac{2 A_{l} ({\rm cos}(\phi/2) + {\rm cos}(k_{\rm saw} L + \phi/2))}{L}, \nonumber\\
S_{zy}(z,t)& = & 2 A_t (\frac{6 z^2}{L^3}-\frac{6 z}{L^2})({\rm
cos}(\phi/2) - \nonumber\\
 & & {\rm cos}(k_{\rm saw} L + \phi/2)),
\label{Equ3}
\end{eqnarray}
where $A_{t}$ resp.~$A_{l}$ are the transversal respectively longitudinal components of the SAW motion, and $L$ is the length of the beam.
The relation between the strain and the electric field is found in the piezoelectric constitutive relation $D_{z} =
\epsilon_{\rm GaAs} E_{z} + e_{z4} S_{zy}$, where $D_z$ is the
electric displacement in $z$ direction, $\epsilon_{\rm GaAs}$ is the dielectric constant of GaAs, and $e_{z4}$ is the appropriate
constant of the piezoelectric tensor, which has to be taken in the [110] coordinate system. This can be solved for $E_{z}$ to give
\begin{equation}
E_{z}(z,t)=-\frac{e_{z4}}{\epsilon_{\rm GaAs}} S_{zy}, \label{Equ4}
\end{equation}
where the contribution of $D_z$ was neglected. This is justified if $1 \ll \sigma_0 / \epsilon_{\rm GaAs} \omega \ll 1 /\lambda_D
k_{\rm saw}$, where $\lambda_D$ is the Debye length~\cite{Ilisavskii:1}. In our case $\sigma_0 / \epsilon_{\rm
GaAs} \omega ~\sim 10^6$ and the condition for disregarding $D_z$ is fulfilled. From Eq.~(\ref{Equ1}-\ref{Equ4}) the current density due
to the anomalous acoustoelectric effect in the center of the beam $z=L/2$ calculates to
\begin{equation}
j_{an}(L/2)=-\frac{\sigma_0 e_{z4}}{2 \ \epsilon_{\rm GaAs}}
\big(\Pi_{zzzy} S_{zy}(L/2)^2 \big), \label{Equ5}
\end{equation}
where the mixed term $S_{zz}, \ S_{xy}$ in Eq.~(\ref{Equ1}) equals to 0, when integrated over one SAW period as the longitudinal and
transversal SAW component are $\pi/2$ phase shifted in time. 
The calculated current in dependence on $\phi$ is shown in the inset
of Fig.~\ref{figure3}(b). Eq.~\ref{Equ5} predicts a square dependence of the anomalous acoustoelectric current on the SAW
amplitude. Using the measured input impedances of the two IDTs it is possible to convert the RF power applied at the transducers
into an estimate for the amplitude of the excited SAW~\cite{Beil:2}. 
Finally, in Fig.~\ref{figure3}(b) shows the maximum of the anomalous
acoustoelectric current component plotted vs. applied SAW amplitude. The curve follows the predicted behavior for not too
small SAW amplitudes, which can be evaluated by fitting this section [cf. Fig.~\ref{figure3}(b)]. The exponent
we extract from this fit evaluates to $2.06 \pm 0.3$, which is in very good agreement with the expected value. From this fit we further
estimate the size of $\Pi_{zzzy}$ to be on the order of $10^5$, which is {\it one order of magnitude larger} than observed for thin
manganite films~\cite{Ilisavskii:1}. The minor deviations of the induced current for small SAW amplitudes suggests that the linear
dependence of conductivity on strain might be an over-simplification. This can be due to different contributions like
dislocations~\cite{Beil:2} or band deformation~\cite{Band:1}.

{\it Conclusion--}
In summary we have shown that shock waves can be generated in nanomechanical resonators. The resonators
contain an electron gas, which we apply to probe the shock waves via the acoustoelectric effect. 
This observation is explained by the presence of two current components, the normal and the anomalous current. 
The anomalous acoustoelectric current is mediated via the strain modulated conductivity in the suspended 2DEG and is described by model calculations. The results will have wide
spread use for quantum electromechanics (QEM), as well as for applications in sensor and communication electronics.
%

{\it Acknowledgements--} We thank J.P. Kotthaus and M.L. Roukes for helpful discussions and support. 
We acknowledge support in part by the Deutsche Forschungsgemeinschaft under contract number Bl-487/3, 
the Air Force OSR (F49620-03-1-0420), the Graduate School of the University of Wisconsin,
and the National Science Foundation (NSF-NIRT). 


\newpage 

\begin{figure}
\caption{Top and side view of the sample geometry and experimental setup. The top part contains a micrograph of the sample, which shows the suspended specimen placed in the delayline formed by two IDTs. Inset gives a scanning electron microscope picture
of the device. The electron gas is
confined in the 100 nm thin membrane. The freely suspended 2DEG interacts with the travelling SAW
via the elliptic motion of clamping points. This mechanical excitation induces acoustoelectric currents measured in the direct current. }
\label{figure0}
\end{figure}

\begin{figure}
\caption{(a) Characterization of the suspended two-dimensional electron gas:  longitudinal magnetoresistance measured in a two point geometry with the  Shubnikov-de-Haas oscillations well pronounced. 
(b) Acoustoelectric current at $T = $~1.5~K with surface acoustic waves driving the current 
from the left (black) and right (red). 
The current is measured with a source drain voltage $V_{\rm DS}$ of 50~$\mu$V. At
$f=f_{\rm saw}$, the IDTs generate acoustical waves and the acoustoelectric current is observed as a dip
in the current. 
The electron gas was tuned into high resistance by applying a backgate voltage ($V_{\rm bg} \approx$
200~V).
Driving an acoustic excitation from the right 
reverses the sign of the normal acoustoelectrical current $I_{\rm n}$ (traces are given for different power levels from 0~dBm to 10~dBm). The remarkable
feature is the region at the edge of the acoustic bandpass: for both directions of the acoustic excitation the anomalous 
acoustoelectric current $I_{\rm an}$ shows no sign reversal.
} \label{figure1}
\end{figure}

\begin{figure}
\caption{
(a) Standing wave measurements -- nonlinear response:
 the current sensitive to the relative phase shift $\phi$ is induced by the acoustic standing wave proportional only to the 
anomalous acoustoelectric current. 
Shown is the phase shift $\phi$ between driving signal applied at
IDTs left and right for increasing acoustic power. The maximum current appears at $180$~deg phase
shift. The applied RF power at the IDTs increases from $-40$~dbm to $-10$~dBm.
(b)  The sinusoidal shape evolves into a $(sin x)/x$-shape, indicating the shock front. At even larger powers the resonator jumps into the 
next higher acoustic mode and the sinusoidal wave form reemerges.
            } \label{figure2}
\end{figure}

\begin{figure}
\caption{
(a) Full scan of the standing wave measurements:  
 the current sensitive to relative phase shift $\phi$ is induced by the acoustic standing wave proportional only to the 
anomalous acoustoelectric current. 
 At large powers the resonator jumps into the 
next higher acoustic mode and the sinusoidal wave form reemerges.
(b) Maximum current vs. applied SAW amplitude and comparison to model calculations:  a square dependence on SAW amplitude is observed,
whereas for small amplitudes deviations are found. Inset shows the calculated dependence of the current on the phase (see
text for details). 
           } \label{figure3}
\end{figure}

\newpage
\begin{figure}[!p]
\vspace{2cm}
\includegraphics[width=12.0cm]{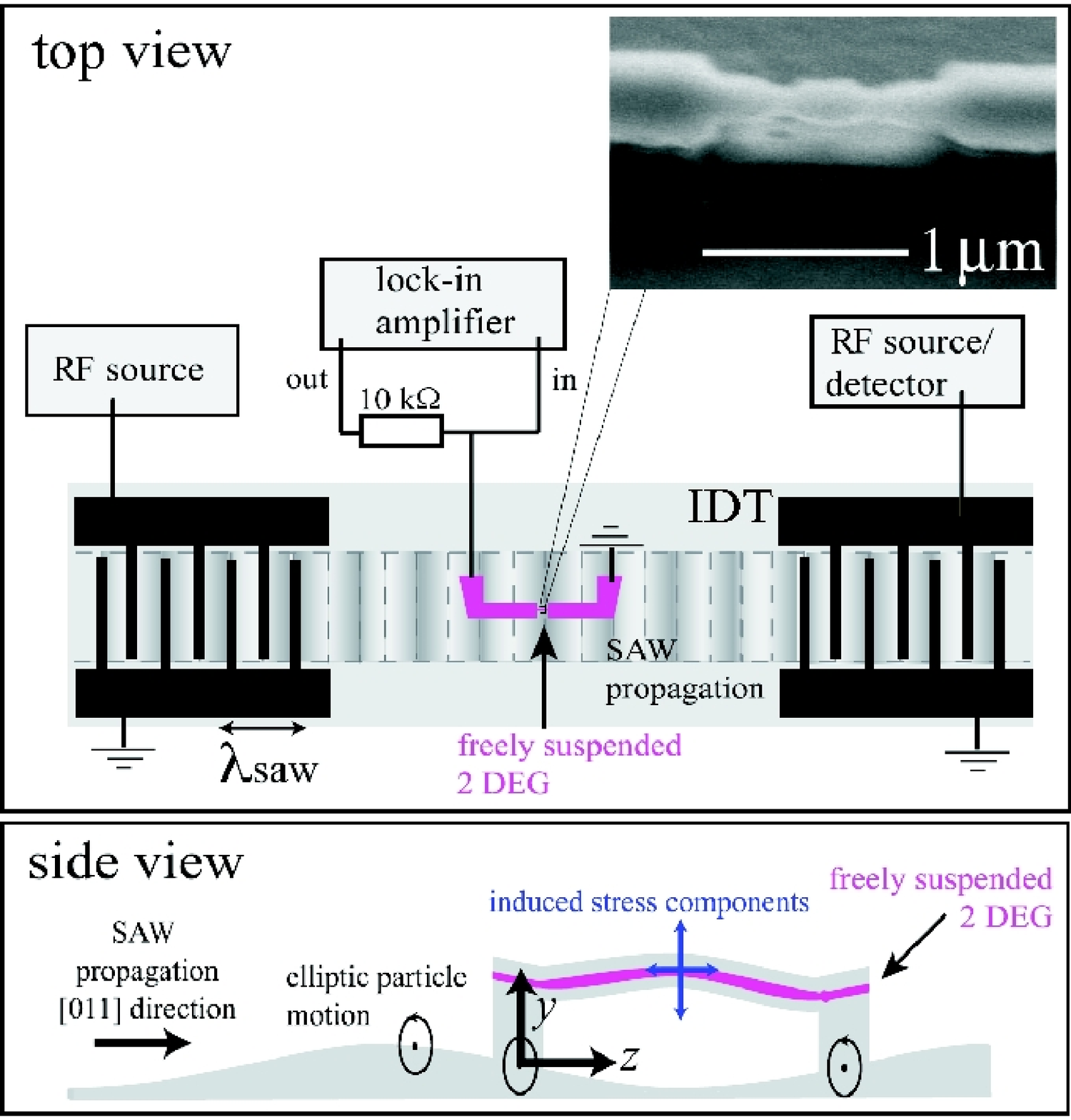}
\center{Beil {\it et al}, Figure 1/4} 
\end{figure}

\begin{figure}[!p]
\vspace{2cm}
\includegraphics[width=12.0cm]{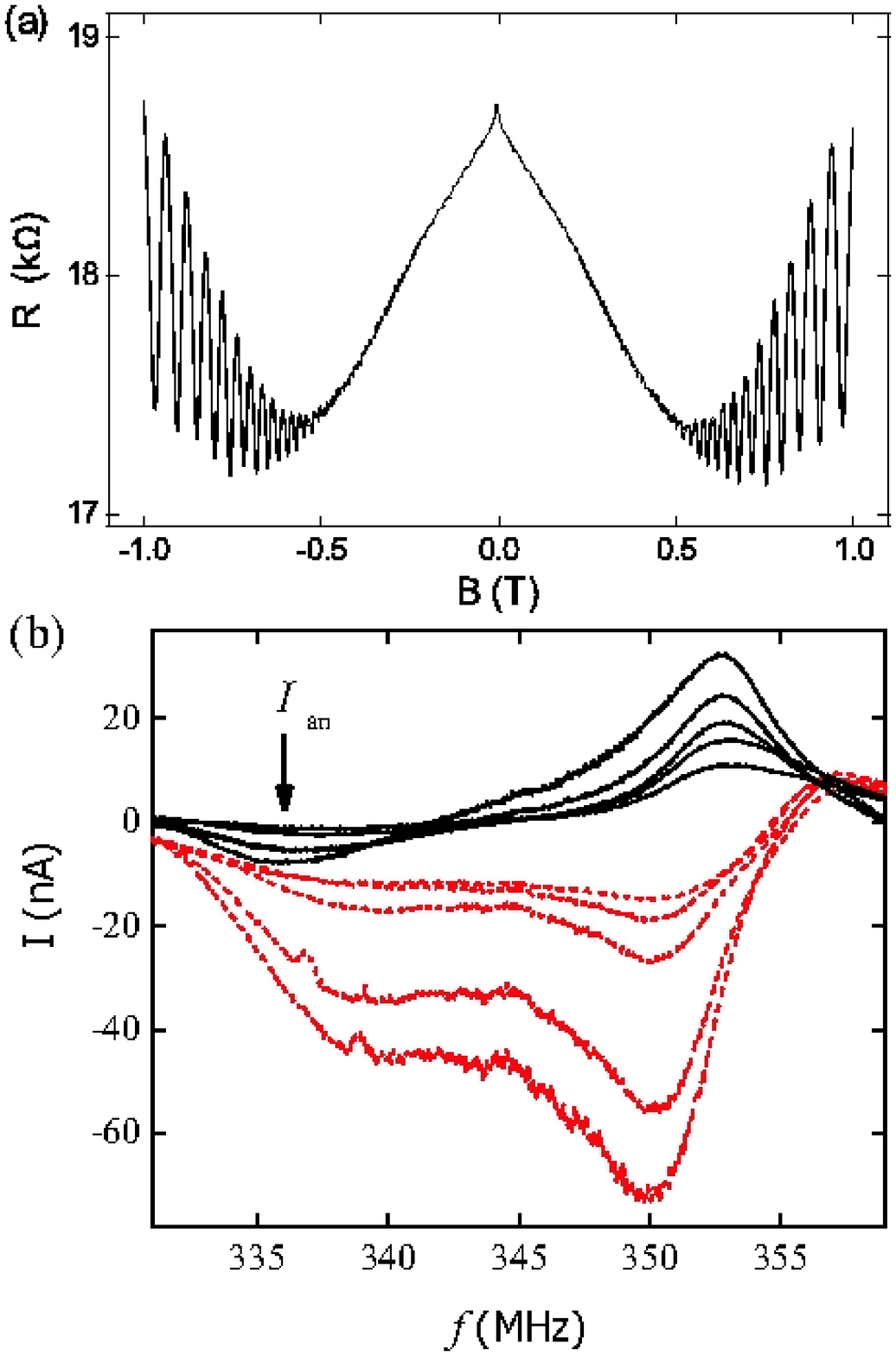}
\center{Beil {\it et al}, Figure 2/4} 
\end{figure}

\begin{figure}[!p]
\vspace{2cm}
\includegraphics[width=12.0cm]{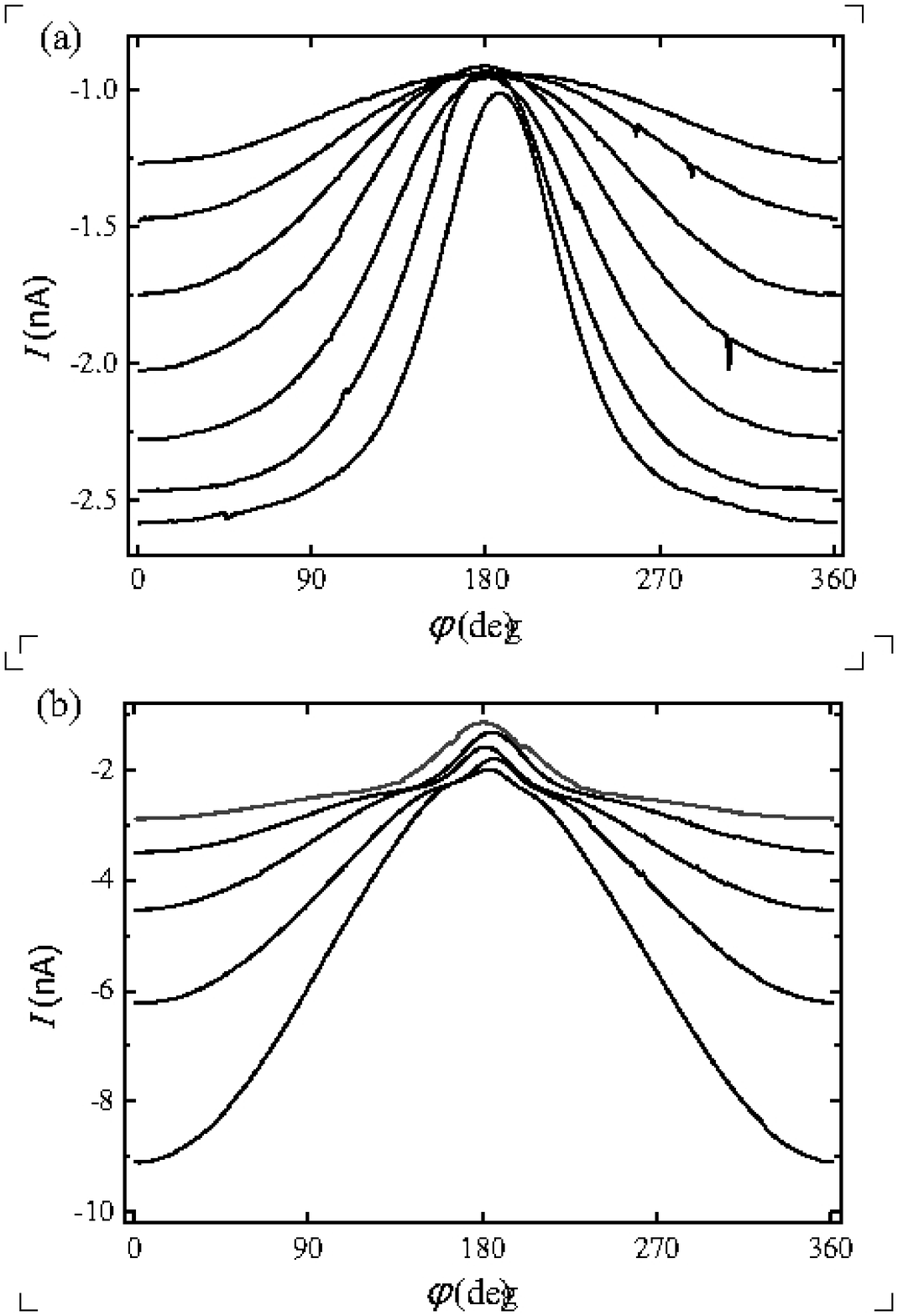}
\center{Beil {\it et al}, Figure 3/4} 
\end{figure}

\begin{figure}[!p]
\vspace{2cm}
\includegraphics[width=12.0cm]{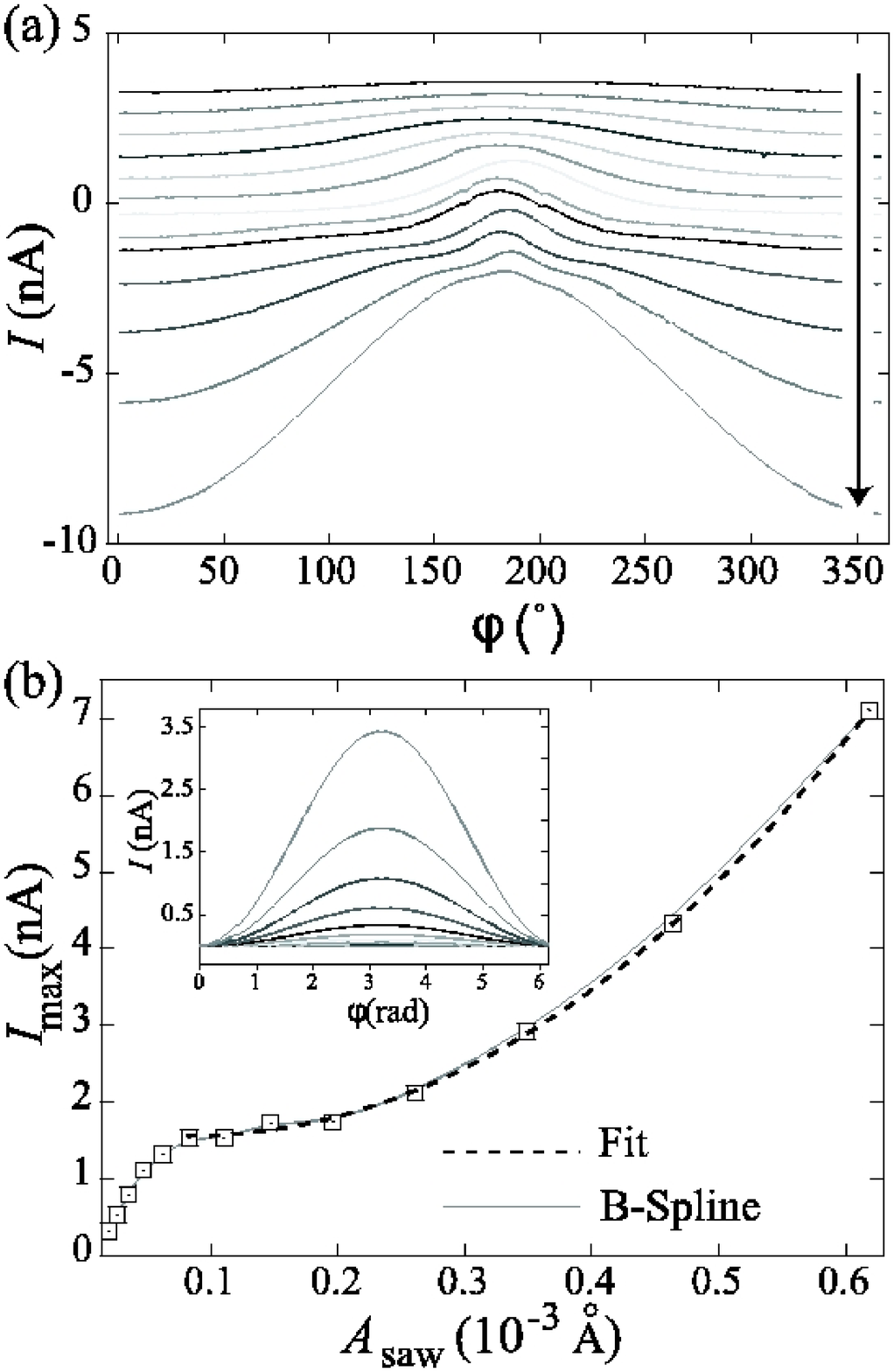}
\center{Beil {\it et al}, Figure 4/4} 
\end{figure}

\end{document}